\documentclass[aps,prd,showpacs,preprintnumbers]{revtex4-2}
\usepackage{amsmath}
\usepackage{subfig}
\usepackage{graphicx}
\usepackage{epstopdf}
\usepackage{float}
\usepackage{subfloat}
\usepackage{amssymb,amsfonts}
\usepackage{latexsym}
\usepackage{epsfig}
\usepackage{amssymb}
\usepackage{color}
\usepackage{bm}
\newcommand{\eq}{\begin{equation}}
	\newcommand{\eeq}{\end{equation}}
\newcommand{\eqn}{\begin{eqnarray}}
	\newcommand{\eeqn}{\end{eqnarray}}

\begin{document}
	\title{Holographic QCD model for $N_f=4$}
	\author{Yidian Chen$^{1}$ }
	\thanks{chenyidian@ucas.ac.cn}
	\author{Mei Huang$^{1}$}
	\thanks{huangmei@ucas.ac.cn}
	\affiliation{$^{1}$ School of Nuclear Science and Technology, University of Chinese Academy of Sciences, Beijing 100049, China}	
	
	\begin{abstract}
		We establish a holographic QCD model for four flavors, where a light scalar field X and a heavy scalar field H are introduced, separately. The H field is responsible for the breaking of $SU(N_f=4)$ to $SU(N_f=3)$.  The ground state and its Regge excitation of meson spectra in the light flavor sector and heavy flavor sector as well as the ligh-heavy mesons are well in agreement with the Particle data group (PDG). Due to the additional introduction of the $H$ field in the model, different Regge slopes for light and heavy mesons can be achieved. 
	\end{abstract}
	
	\maketitle
	
	\section{Introduction}\label{sec-1}
	
	There are six flavors of quarks as fundamental building blocks of matter, $u,d,s$ in light flavor sector with current quark mass ranging from several ${\rm MeV}$ to hundred ${\rm MeV}$, and $c,b,t$ in heavy flavor sector with current mass $1.5 -172 {\rm GeV}$, which is much larger than the QCD scale $\Lambda_{QCD}$, thus the flavor symmetry is badly broken. The chiral symmetry and its spontaneous breaking is dominant in light flavor sector while the heavy flavor sector is characterized by heavy-quark symmetry \cite{Isgur:1991wq}. Apart from the puzzles of flavor hierarchy and flavor desert issues \cite{Xing:2020ijf}, the heavy flavor sector especially hadrons composed of charm qaurk attracts much attention both in particle physics and heavy-ion physics. Recently, the Belle collaboration and the BESIII collaboration have reported the observations of multiquark exotics with heavy flavor,see review article \cite{Brambilla:2019esw}. On the other hand, charmonium suppression was suggested as a signal of deconfinement phase transition almost 35 years ago \cite{Matsui:1986dk}, and has been always an important topic in heavy-ion collisions.
	
	Theoretically, because the large mass of heavy quark, multi energy scales from high-energy perturbative to low energy nonperturbative contributions are involved in the calculations, the description of hadrons containing one or two heavy quarks is rather challenging. A nonrelativistic treatment of the heavy quarkonium dynamics has been developed in nonrelativistic QCD (NRQCD),  and the heavy quark effective theory (HQET) has been used to describe systems with only one heavy quark, see review article \cite{QuarkoniumWorkingGroup:2004kpm}. Also an extended linear sigma model for four quark flavors has been developed in Ref.\cite{Eshraim:2014eka}, where light flavor mesons,light-heavy flavor mesons as well as charmonium can be described reasonably well.
	
	In recent decades, the gravity/gauge duality, or anti-de Sitter/conformal field theory (AdS/CFT) correspondence \cite{Maldacena:1997re,Gubser:1998bc,Witten:1998qj} offers a new possibility to tackle the difficulty of strongly coupled gauge theories, for reviews see Refs.\cite{Reviews} .The ‘bottom-up’ holographic models of QCD based on AdS/CFT have emerged as an effective approach to the low energy phenomenology of QCD,  which have been widely and successfully used in describing hadron physics, especially light flavor hadron spectra \cite{Erlich:2005qh,Karch:2006pv,mesons,DHQCD,VHQCD}. On the other hand, the heavy flavor hadron spectra has been seldomly investigated in the framework of holographic QCD till very recently \cite{Bai:2013rza,Liu:2016iqo,Liu:2016urz,Ballon-Bayona:2017bwk,Momeni:2020bmy}. 
	
	The hard-wall holographic QCD model has been directly extended to four flavors in Refs.\cite{Ballon-Bayona:2017bwk,Momeni:2020bmy}, which in some sense can be regarded as the 5D version of the extended linear sigma model for four quark flavors\cite{Eshraim:2014eka}, where the ground states of light flavor mesons,light-heavy flavor mesons as well as charmonium can be described. A holographic model for heavy-light mesons has been discussed in \cite{Bai:2013rza}. Holographic heavy-light chiral effective action in D4-D8 system has been constructed in \cite{Liu:2016iqo,Liu:2016urz}.
	
	In this work, we are going to establish a holographic QCD model for four flavors $N_f=4$, where the ground states and the higher excitation states of light flavor mesons, light-heavy flavor mesons as well as charmonium can be generated. Moreover, different Regge slopes for light flavor spectra and heavy flavor meson spectra can be realized in this framework. The paper is organized as following: After Introduction, in Sec.\ref{sec-2} we introduce the framework of the $N_f=4$ holographic model; Then in Sec.\ref{sec-3} and \ref{sec-4} we introduce how to calculate the meson spectra and decay constant, and we show our numerical results in Sec.\ref{sec-5}, and at last we give conclusion in Sec.\ref{sec-conclusion}.
	
	\section{5D Model Setup}\label{sec-2}
	
	The QCD can be described as a holographic 5-dimensional model according to AdS/CFT duality.
	In this section, we introduce the holographic QCD model, which includes the complex scalar fields $X$, two gauge fields $L_M^a$ and $R_M^a$, which correspond to the $\langle\bar{q}_Rq_L\rangle$ and $\langle\bar{q}_{L,R}\gamma_\mu t^aq_{L,R}\rangle$ operators from boundary theory, respectively\cite{Erlich:2005qh}. 
	Where $t^a,~a=1,2,...N_f^2-1$ are the generators of the $SU(N_f)$ group. 
	In this paper, we take $N_f=4$ and let the generators satisfy ${\rm Tr}(t^at^b)=\frac{1}{2}\delta^{ab}$. 
	The scalar field $X$ describes the breaking of the chiral symmetry of light flavor quarks in QCD, at the UV boundary, which has $X(z)|_{z\to z_{UV}}\to M_q z+\sigma z^3$. 
	It can be seen that the first term describes the explicit breaking of the chiral symmetry due to the non-zero quark mass, and the second term describes the chiral condensation of QCD. 
	Unlike the traditional soft wall and hard wall models, the dilaton field and hard cutoff $z_m$ are introduced simultaneously in this model.
	With reference to \cite{Karch:2006pv}, we introduce the dilaton field $\phi=\mu^2z^2$ to implement the linear Regge behavior of the light hadrons, which contains information about the gluon dynamics. Here $\mu^2$ is proportional to the slope of Regge trajectories of light flavor mesons. The reason for the introduction of hard cutoff is to realize the difference in Regge slope of light and heavy mesons, which has almost no effect on light mesons, similar to the soft wall model, while for heavy mesons, the effect is similar to the hard wall model due to the heavy quark mass.
	
	In addition to this, another scalar field $H$ is introduced to describe the difference between the light and heavy quark masses. Since the scalar field $X$ breaks the $SU(N_f)_L\times SU(N_f)_R\simeq SU(N_f)_V\times SU(N_f)_A$ symmetry of the system to $SU(N_f)_V$, the residual symmetry makes the equations of motion of the light vector mesons indistinguishable from those of the charmonium. Therefore, the $H$ field is introduced to describe the difference between the $\rho$ mesons and the $J/\psi$ mesons. 
	The $H$ field acts similarly to the $\Psi$ field of the D4-D8 model\cite{Liu:2016iqo,Liu:2016urz} and arises from the large difference between the light and heavy quark masses. From the top-down point of view, this difference is described by the distance of the light and heavy flavor brane, thus introducing the $\Psi$ field in the DBI action. 
	Equivalently, by introducing the $H$ field, we break the residual $SU_V(N_f)$ symmetry to $SU_V(N_f-1)$.
	Therefore, the suitable configuration for the auxiliary $H$ field can break $SU(4)_V$ symmetry to $SU(3)_V$ in the case of $N_f = 4$.  
	
	Therefore, the total 5D action is
	\begin{eqnarray}\label{action}
		S_M &=& -\int_\epsilon^{z_m} d^5 x \sqrt{-g} ~ e^{-\phi} ~{\rm Tr} \Big \{ (D^M X)^{\dag} (D_M X)  + m_5^2 |X|^2 \cr
		&+& \frac{1}{4 g_5^2} \left ( L^{MN} L_{MN} + R^{MN} R_{MN} \right )+(D^M H)^{\dag} (D_M H)  + m_5^2 |H|^2 
		\Big \} \, ,
	\end{eqnarray}
	where $D_{M} X=\partial_M X -iL_M X +i X R_M$ and $D_M H = \partial_M H - i V_M^{15} H - i H V_M^{15}$ are the
	covariant derivative of the scalar field $X$ and $H$, respectively. 
	According to \cite{Erlich:2005qh}, the coupling constant $g_5^2=12\pi^2/N_c$. 
	The 5-dimensional masses of the scalar fields $H$ and $X$ are fixed to $m_5^2=\Delta(\Delta-4)=-3$, due to the conformal dimension of the operator $\langle\bar{q}q\rangle$. 
	The strength of the non-abelian gauge field $L_M$ and $R_M$ are defined as
	\begin{eqnarray}
		L_{MN} &=& \partial_M L_N - \partial_N L_M - i \left [ L_M , L_N \right ], \nonumber\\
		R_{MN} &=& \partial_M R_N - \partial_N R_M - i \left [ R_M , R_N \right ],
	\end{eqnarray}
	with $L_M=L_M^at^a$ and $R_M=R_M^at^a$. it is convenient to rewrite the left and right gauge fields in terms of vector and axial fields , i.e. $L_M = V_M + A_M$ and $R_M = V_M-A_M$. 
	The bilinear field $X$ can be decomposed as
	\begin{eqnarray}
		X = e^{i\pi^at^a} \, X_0 \, e^{i\pi^bt^b},
	\end{eqnarray}
	where $X_0={\rm diag}[v_l(z),v_l(z),v_s(z),v_c(z)]$. 
	Similar to the case of $N_f = 2$, the $v_{l,s,c}(z)$ field has the behavior of $v_{l,s,c}(z)\to M_{l,s,c}z+\Sigma_{l,s,c}z^3$ at the UV boundary.
	The auxiliary field $H$ is given as $H={\rm diag}[0,0,0,h_c(z)]$. 
	Since the $H$ field reflects the effect of the charm quark mass, the behavior of $h_c$ at the UV boundary has $h_c(z)\to m_c z$. 
	It is worth noting that we consider $m_c\neq M_c$, in order to obtain better results of the vector meson mass spectra.
	
	In this paper, we do not consider the back reaction of the flavor-brane, so the Poincare patch of the 5-dimensional AdS spacetime is
	\begin{eqnarray}\label{eq.1}
		ds^2=\frac{L^2}{z^2}\left(dz^2+\eta_{\mu\nu}dx^{\mu}\,dx^{\nu}\right),
	\end{eqnarray}
	with $\mu,\nu=0,1,2,3$ and $\eta_{\mu\nu}=\rm{diag}[-1,1,1,1]$. 
	Without loss of generality, the AdS radius $L$ is set to 1. 
	According to the AdS/CFT duality, the fifth dimensional coordinate $z$ describes the running of the energy scale, $z\to 0$ corresponds to UV boundary and $z\to z_m$ corresponds to IR boundary.
	
	\subsection{Expansion of the Action}
	
	To obtain the meson masses as well as the three and four point coupling constants, the action Eq.(\ref{action}) is expanded to quartic order, which is given as
	\begin{eqnarray}
		S=S^{(0)}+S^{(2)}+S^{(3)}+S^{(4)},\nonumber
	\end{eqnarray}
	where
	\begin{eqnarray}
		S^{(0)}&=&-\int_0^{z_m} d^5x~\Bigg\{\frac{e^{-\phi(z)}}{z^3}\bigg(2v_l^\prime(z) v_l^\prime(z)+v_s^\prime(z)v_s^\prime(z)+v_c^\prime(z)v_c^\prime(z)\bigg)+\frac{e^{-\phi(z)}}{z^5}m_5^2\bigg(2v_l(z)^2+v_s(z)^2+v_c(z)^2\bigg)\nonumber\\
		&&+\frac{e^{-\phi(z)}}{z^3}\bigg(h_c^\prime(z)h_c^\prime(z)\bigg)+\frac{e^{-\phi(z)}}{z^5}m_5^2h_c(z)^2\Bigg\},\label{eom-vev}\\
		S^{(2)}&=&-\int_0^{z_m} d^5x\Bigg\{\eta^{mn}\frac{e^{-\phi(z)}}{z^3}\bigg((\partial_m\pi^a-A_m^a)(\partial_n\pi^b-A_n^b)M_A^{ab}-V_m^aV_n^bM_V^{ab}+V_m^{15}V_n^{15}m_V^{15,15}\bigg)\nonumber\\
		&&+\frac{e^{-\phi(z)}}{4g_5^2z}\eta^{mp}\eta^{nq}\bigg(V_{mn}V_{pq}+A_{mn}A_{pq}\bigg)\Bigg\},\label{eom-2}\\
		S^{(3)}&=&-\int_0^{z_m} d^5x\Bigg\{\eta^{mn}\frac{e^{-\phi(z)}}{z^3}\bigg(2(A_m^a-\partial_m\pi^a)V_n^b\pi^cg^{abc}+V_m^a(\partial_n(\pi^b\pi^c)-2A_n^b\pi^c)h^{abc}\bigg)\nonumber\\
		&&+\frac{e^{-\phi(z)}}{2g_5^2z}\eta^{mp}\eta^{nq}(V_{mn}^aV_p^bV_q^c+V_{mn}^aA_p^bA_q^c+A_{mn}^aV_p^bA_q^c+A_{mn}^aA_p^bV_q^c)f^{bca}\Bigg\},\label{eom-3}\\
		S^{(4)}&=&-\int_0^{z_m} d^5x\Bigg\{\eta^{mn}\frac{e^{-\phi(z)}}{z^3}\bigg([\partial_m\pi^a-A_m^a][A_n^b\pi^c\pi^d-\frac{1}{3}\partial_n(\pi^b\pi^c\pi^d)]l^{abcd}+V_m^aV_n^b\pi^c\pi^d(h^{abcd}-g^{acbd})\nonumber\\
		&&+[\frac{1}{2}\partial_m(\pi^a\pi^b)-A_m^a\pi^b][\frac{1}{2}\partial_n(\pi^c\pi^d)-A_n^c\pi^d]k^{abcd}\bigg)+\frac{e^{-\phi(z)}}{4g_5^2z}\eta^{mp}\eta^{nq}(V_m^aV_n^bV_p^cV_q^d+A_m^aA_n^bV_p^cV_q^d+V_m^aV_n^bA_p^cA_q^d\nonumber\\
		&&+A_m^aA_n^bA_p^cA_q^d+2V_m^aA_n^bV_p^cA_q^d+2A_m^aV_n^bV_p^cA_q^d)f^{abcd}\Bigg\}.\label{eom-4}
	\end{eqnarray}
	It is worth noting that $\eta^{mn}$ is the 5-dimensional Minkowski spacetime metric and $V(A)_{mn}=\partial_{m}V(A)_{n}-\partial_{n}V(A)_{m}$ is the abelian field strengths. 
	Here, $M_A^{ab}$, $M_V^{ab}$, $m_V^{ab}$, $h^{abc}$, $g^{abc}$, $g^{abcd}$, $l^{abcd}$, $h^{abcd}$, $k^{abcd}$, and $f^{abcd}$ are defined as
	\begin{eqnarray}
		&M_A^{ab}={\rm Tr}(\{t^a,X_0\}\{t^b,X_0\}),\qquad &M_V^{ab}={\rm Tr}([t^a,X_0][t^b,X_0]),\nonumber\\
		&m_V^{15,15}={\rm Tr}(\{H,t^{15}\}\{H,t^{15}\}),\qquad &f^{abcd}=f^{\alpha ab}f^{\alpha cd}\nonumber\\
		&h^{abc}=i~{\rm Tr}([t^a,X_0]\{t^b,\{t^c,X_0\}),\qquad &g^{abc}=i~{\rm Tr}(\{t^a,X_0\}[t^b,\{t^c,X_0\}]),\nonumber\\
		&h^{abcd}={\rm Tr}([t^a,X_0][t^b,\{t^c,\{t^d,X_0\}\}]),\qquad &g^{abcd}={\rm Tr}([t^a,\{t^b,X_0\}][t^c,\{t^d,X_0\}]),\nonumber\\
		&l^{abcd}={\rm Tr}(\{t^a,X_0\}\{t^b,\{t^c,\{t^d,X_0\}),\qquad &k^{abcd}={\rm Tr}(\{t^a,\{t^b,X_0\}\{t^c,\{t^d,X_0\}),
	\end{eqnarray}
	with structure constant $f^{abc}$ of $SU(4)$ Lie algebra. 
	
	The vectors, axial vectors and pseudoscalars mesons can be described by $4\times 4$ matrices:
	\begin{eqnarray}
		V &=&  V^a t^a  = \frac{1}{\sqrt 2}
		\left ( \begin{matrix}
			\frac{\rho^0}{\sqrt{2}} + \frac{\omega'}{\sqrt{6}} + \frac{\psi}{\sqrt{12}}  &  \rho^{_{+}}  &  K^{_{*+}}  &  \bar D^{_{*0}}  \\
			\rho^{_{-}}   & -\frac{\rho^0}{\sqrt{2}}  + \frac{\omega'}{\sqrt{6}}  + \frac{\psi}{\sqrt{12}}   &  K^{_{*0}}  &  D^{_{*-}}  \\
			K^{_{*-}}  &  \bar K^{_{*0}}  & - \sqrt{\frac23} \omega' + \frac{\psi}{\sqrt{12}}  &  D_s^{_{*-}}  \\
			D^{_{*0}} &  D^{_{*+}}  &  D_s^{_{*+}}  & - \frac{3}{\sqrt{12}} \psi
		\end{matrix} \right ), \\
		A &=&  A^a t^a  =  \frac{1}{\sqrt 2} \left(\begin{array}{cccc}
			\frac{a^0_1}{\sqrt{2}}  + \frac{f_1}{\sqrt{6}}+\frac{\chi_{c1}}{\sqrt{12}}
			& a^{_{+}}_1 & K_{1}^{_{+}}& \bar {D}_{1}^{_{0}} \\
			a_1^{_{-}}&- \frac{ a^{_{0}}_1}{\sqrt{2}} +\frac{f_1}{\sqrt{6}}+\frac{\chi_{c1}}{\sqrt{12}}
			& K_{1}^{_{0}}& D_{1}^{_{-}}\\
			K_{1}^{_{-}}& \bar{K}_{1}^{_{0}}  & -\sqrt{\frac23}(f_1)+\frac{\chi_{c1}}{\sqrt{12}}& D_{s1}^{_{-}}\\
			{D}_{1}^{_{0}}&D_{1}^{_{+}}&D_{s1}^{_{+}}&- \frac{3}{\sqrt{12}}\chi_{c1}
		\end{array}
		\right), \\
		\pi &=& \pi^a t^a = \frac{1}{\sqrt 2}
		\left (\begin{matrix}
			\frac{\pi^{_{0}}}{\sqrt{2}}  + \frac{\eta}{\sqrt{6}}  + \frac{\eta_c}{\sqrt{12}}  &  \pi^{_{+}}  &  K^{_{+}}  &  \bar D^{_{0}}  \\
			\pi^{_{-}}   & -\frac{\pi^0}{\sqrt{2}}  + \frac{\eta}{\sqrt{6}}  + \frac{\eta_c}{\sqrt{12}}   &  K^{_{0}}  &  D^{_{-}} \\
			K^{_{-}}  &  \bar K^{_{0}}  & - \sqrt{\frac23} \eta + \frac{\eta_c}{\sqrt{12}}  &  D_s^{_{-}}  \\
			D^{_{0}} &  D^{_{+}}  &  D_s^{_{+}}  & - \frac{3}{\sqrt{12}} \eta_c
		\end{matrix} \right ).
	\end{eqnarray}
	By substituting the above matrices into the action Eqs.(\ref{eom-2},\ref{eom-3},\ref{eom-4}), the wave functions, mass spectra, decay constants and coupling constants of the mesons can be obtained.
	
	\subsection{Scalar Vacuum Expectation Value}
	
	The equations of motion(EOMs) of scalar vacuum expectation value $v_{l,s,c}$ can be obtained by Eq.(\ref{eom-vev})
	\begin{eqnarray}
		-\frac{z^3}{e^{-\phi(z)}}\partial_z \frac{e^{-\phi}}{z^3}\partial_z v_q(z)+\frac{m_5^2}{z^2}v_q(z)=0.
	\end{eqnarray}
	The analytic solutions are
	\begin{eqnarray}
		v_q(z)=C_1(q)~z~\sqrt{\pi}~U(\frac{1}{2},0,\phi)-C_2(q)~z~L(-\frac{1}{2},-1,\phi),
	\end{eqnarray}
	where $U$ is confluent hypergeometric function and $L$ is the generalized Laguerre polynomial. 
	Here, the $q$ following the constants $C_1$ and $C_2$ indicate that for different quarks, the values are different.
	Expanding $v_q$ at the UV boundary gives
	\begin{eqnarray}
		v_q(z)|_{z\to 0}=2~C_1(q)~z+\bigg( C_2(q)~\mu^2+C_1(q)\Big[-\mu^2+2\gamma_E\mu^2+2\mu^2{\rm Log}~z+2\mu^2{\rm Log}~\mu+\mu^2\Psi(\frac{3}{2})\Big]\bigg)z^3,
	\end{eqnarray}
	with the Euler's constant $\gamma_E$ and the digamma function $\Psi$.
	It can be seen that the quark mass $M_q$ is related to $C_1(q)$, while the quark condensation $\Sigma_q$ is related to $C_1(q)$ and $C_2(q)$. Referring to Ref.\cite{DHQCD}, $C_2(q)$ will produce the nonlinear spectra of $a_1$ mesons, so we set $C_2(q) = 0$. For the auxiliary field $h_c$, there is similar result:
	\begin{eqnarray}
		h_c(z)=D_1~z~\sqrt{\pi}~U(\frac{1}{2},0,\phi)-D_2~z~L(-\frac{1}{2},-1,\phi).
	\end{eqnarray}
	As we discussed above, we set $D_2 = 0$. 
	This means that the difference between the light-flavored and heavy-flavored vector mesons masses comes from the mass term of the heavy quark. 
	
	\section{wave functions, mesons spectra and decay constants}\label{sec-3}
	
	\subsection{Wave functions and mesons spectra}
	
	The EOMs of the transverse part of vector fields are obtained by Eq.(\ref{eom-2}):
	\begin{eqnarray}\label{eomv}
		\left(-\frac{z}{e^{-\phi}}\partial_z\frac{e^{-\phi}}{z}\partial_z+\frac{2g_5^2(m_V^{ab}-M_V^{ab})}{z^2}\right) V^a_{\mu\perp}(q,z)=-q^2V^a_{\mu\perp}(q,z),
	\end{eqnarray}
	where $V^{z,a} = 0$ gauge is considered and $V_\perp^{\mu,a}$ satisfy $\partial_\mu V_\perp^{\mu,a}=0$. 
	And $V^a_{\mu\perp}(q,z)$ are the 4D Fourier transform of $V^a_{\mu\perp}(x,z)=\int d^4q e^{iqx}V^a_{\mu\perp}(q,z)$. 
	According to AdS/CFT duality, the fields $V^a_{\mu\perp}(q,z)$ can be written as $V^a_{\mu\perp}(q,z)=V^{(n)a}_{\mu\perp}(q,z)+V^{0a}_{\mu\perp}V_{\perp}^a(q,z)$, where $V_{\perp}^a(q,z)$ are bulk-to-boundary propagators and $V^{0a}_{\mu\perp}$ respect to the source. 
	The discrete mass spectra and wave functions can be obtained by setting KK towers $V^{(n)a}_{\mu\perp}(q,z)$ with boundary condition $V^{(n)a}_{\mu\perp}(z)|_{q,z\to0}=0$ and $\partial_zV^{(n)a}_{\mu\perp}(z)|_{z\to z_m}= 0$. 
	
	Similar to the vector fields, the EOMs of the transverse part of axial vector fields are obtained by Eq.(\ref{eom-2}):
	\begin{eqnarray}\label{eoma}
		\left(-\frac{z}{e^{-\phi}}\partial_z\frac{e^{-\phi}}{z}\partial_z+\frac{2g_5^2M_A^{ab}}{z^2}\right) A^a_{\mu\perp}(q,z)=-q^2A^a_{\mu\perp}(q,z),
	\end{eqnarray}
	where $A^{z,a} = 0$ gauge and transverse condition $\partial_\mu A_\perp^{\mu,a}=0$ are considered.  
	And $A^a_{\mu\perp}(q,z)$ are the 4D Fourier transform of $A^a_{\mu\perp}(x,z)=\int d^4q e^{iqx}A^a_{\mu\perp}(q,z)$. 
	The fields $A^a_{\mu\perp}(q,z)$ can also be written as $A^a_{\mu\perp}(q,z)=A^{(n)a}_{\mu\perp}(q,z)+A^{0a}_{\mu\perp}A^a_{\perp}(q,z)$, with bulk-to-boundary propagators $A^a_{\perp}(q,z)$ and KK towers of axial vector mesons $A^{(n)a}_{\mu\perp}(q,z)$. 
	For the axial vector mesons wave functions, the boundary conditions of the axial vector field $A$ are set as $A^{(n)a}_{\mu\perp}(q,z)|_{z\to0}=0$ and $\partial_zA^{(n)a}_{\mu\perp}(q,z)|_{z\to z_m}= 0$. 
	
	The longitudinal part of the axial vector fields and the pseudoscalar fields have mixing, and their EOMs can be described through Eq.(\ref{eom-2}) as
	\begin{eqnarray}	
		&&q^2\partial_z \varphi^a(q,z)+\frac{2g_5^2M_A^{ab}}{z^2}\partial_z \pi^a(q,z)=0\,,\label{eom-pi-1}\\			
		&&\frac{z}{e^{-\phi}}\partial_z \left(\frac{e^{-\phi}}{z}\partial_z\varphi^a(q,z) \right)  -
		\frac{2g_5^2M_A^{ab}}{z^2}\left(\varphi^a(q,z)-\pi^a(q,z)\right)=0\,,\label{eom-pi-2} 					
	\end{eqnarray}
	where $\varphi$ is the longitudinal part of the axial vector fields $A_{\mu\parallel}=\partial_\mu\varphi$. 
	Similarly, the boundary condition of the pseudoscalar fields are set to $\pi^{(n)a}(q,z)|_{z\to0}=\varphi^{(n)a}(q,z)|_{z\to0}=0$ and $\partial_z\varphi^{(n)a}(q,z)|_{z\to z_m}= 0$. 
	
	However, for the lowest mode obtained under the above boundary conditions, numerical calculations show that it is not Goldstone mode, and the lowest eigenvalue does not converge to 0 for quark mass $M_q\to 0$. To obtain Goldstone mode, the quark mass $M_q = 0$ and the eigenvalue $q^2 = 0$ are chosen, and the analytical solution of $\varphi$ is $(e^{\mu^2z^2}-1)/(2\mu^2)/$ given by Eq.(\ref{eom-pi-1},\ref{eom-pi-2}). Obviously, $\varphi$ does not satisfy $\partial_z\varphi^{(n)a}(q,z)|_{z\to z_m}= 0$ at IR boundary. Therefore, the boundary condition of the pseudoscalar meson is set to $\pi^{(n)a}(q,z)|_{z\to0}=\varphi^{(n)a}(q,z)|_{z\to0}=0$ and $\partial_z\varphi^{(n)a}(q,z)|_{z\to z_m}\propto\partial_z\pi^{(n)a}(q,z)|_{z\to z_m}$ for the Goldstone mode.
	
	\subsection{Decay constants}
	
	Similar to Ref.\cite{Erlich:2005qh}, we derive the decay constants of pseudoscalar, vector and axial vector mesons in this section. According to the AdS/CFT correspondence, the vector current two-point function can be obtained by differentiating the on-shell action twice,
	\begin{eqnarray}\label{2-function}
		\int_xe^{iqx}\langle J_{V,\mu}^{a}(x)J_{V,\mu}^{b}(0)\rangle = \delta^{ab}(q_\mu q_\nu-q^2g_{\mu\nu})\Pi_V(q^2),\\
		\Pi_V(q^2)=-\frac{e^{-\phi(z)}}{g_5^2q^2}\frac{\partial_zV(q,z)}{z}|_{z= \epsilon\to 0},
	\end{eqnarray}
	with $V(q,\epsilon)=1$. For the Sturm-Liouville equation Eq.(\ref{eomv}), $V(q, z)$ can be expressed as
	\begin{eqnarray}\label{2-function}
		V(z^\prime)=\frac{e^{-\phi(z)}}{z}\sum_n\frac{\psi_{V^n}^\prime(\epsilon)\psi_{V^n}(z^\prime)}{q^2-m_n^2},
	\end{eqnarray}
	where $\psi_{V^n}$ is the eigenfunction of the vector equation Eq.(\ref{eomv}) and satisfies the normalization condition $\int dz~e^{A(z)-\phi(z)}\\
	\psi_{V^n}\psi_{V^m}=\delta_{nm}$. Taking the above equation to the two-point function yields
	\begin{eqnarray}\label{2-function}
		\Pi_V(q^2)=-\frac{1}{g_5^2q^2}\sum_n\frac{[e^{A(\epsilon)-\phi(\epsilon)}\psi_{V^n}^\prime(\epsilon)]^2}{q^2-m_n^2}.
	\end{eqnarray}
	Considering the definition of the decay constant $\langle0|J_{V}^{\mu}|V(p)\rangle=if_{V}p^{\mu}$, it can be given as
	\begin{eqnarray}
		F_{V^n}^2=\frac{[e^{A(\epsilon)-\phi(\epsilon)}\psi_{V^n}^\prime(\epsilon)]^2}{g_5^2}|_{\epsilon\to 0}.
	\end{eqnarray}
	Similar to the vector, the decay constant of the axial vector meson is
	\begin{eqnarray}
		F_{A^n}^2=\frac{[e^{A(\epsilon)-\phi(\epsilon)}\psi_{A^n}^\prime(\epsilon)]^2}{g_5^2}|_{\epsilon\to 0},
	\end{eqnarray}
	where $\psi_{A^n}$ is the eigenfunction of the axial vector part and satisfies the normalization condition $\int dz~e^{A(z)-\phi(z)}\\
	\psi_{A^n}\psi_{A^m}=\delta_{nm}$.
	
	Since $\Pi_A(q^2)\to-f_\pi^2q^2$ with $q^2\to 0$, the decay constant of the pseudoscalar meson is
	\begin{eqnarray}
		f_{\pi}^2=-\frac{e^{A(\epsilon)-\phi(\epsilon)}\partial_zA(0,\epsilon)}{g_5^2}|_{\epsilon\to 0},
	\end{eqnarray}
	where $A(0,\epsilon)$ is the solution of Eq.(\ref{eoma}) with $q = 0$ and satisfies the boundary condition $A^\prime(0,z_m)=0$ and $A(0,\epsilon)=1$.
	
	\section{three and four point functions}\label{sec-4}
	
	The three-point interaction of mesons can be obtained by the cubic order term of 5D action. 
	From Eq.(\ref{eom-3}), it can be seen that the cubic order of the action can be divided into four parts $S_{VVV}$, $S_{VAA}$, $S_{VA\pi}$ and $S_{V\pi\pi}$, where	
	\begin{eqnarray}
		S_{VVV}&=&-\int_0^{z_m} d^5x\frac{e^{-\phi(z)}}{2g_5^2z}f^{bca}V^{\mu\nu,a}V_\mu^bV_\nu^c,\\
		S_{VAA}&=&-\int_0^{z_m}d^5x\frac{e^{-\phi(z)}}{2g_5^2z}f^{bca}\bigg(V^{\mu\nu,a}A_\mu^bA_\nu^c+2A^{\mu\nu,a}V_\mu^bA_\nu^c\bigg),\\
		S_{VA\pi}&=&-\int_0^{z_m} d^5x\frac{e^{-\phi(z)}}{z^3}2V^{\mu,a}A_\mu^b\pi^c(g^{bac}-h^{abc}),\\
		S_{V\pi\pi}&=&-\int_0^{z_m} d^5x\frac{e^{-\phi(z)}}{z^3}V^{\mu,a}\pi^b\partial_\mu\pi^c(h^{abc}+h^{acb}-2g^{cab}).
	\end{eqnarray}
	
	Substituting the 5D eigenfunctions of the mesons into the action, the coupling constants of three-point interactions are obtained as
	\begin{eqnarray}\label{3cc-1}
		g_{VVV}&=&\int_0^{z_m} dz\frac{e^{-\phi(z)}}{2g_5^2z}f^{bca}\psi_{V^{(n)}}^a\psi_{V^{(m)}}^b\psi_{V^{(k)}}^c,\\
		g_{VAA}&=&\int_0^{z_m}dz\frac{e^{-\phi(z)}}{2g_5^2z}f^{bca}\psi_{V^{(n)}}^a\psi_{A^{(m)}}^b\psi_{A^{(k)}}^c,\label{3cc-2}\\
		g_{VA\pi}&=&\int_0^{z_m} dz\frac{e^{-\phi(z)}}{z^3}2\psi_{V^{(m)}}^a\psi_{A^{(m)}}^b\psi_{\pi^{(k)}}^c(g^{bac}-h^{abc}),\label{3cc-3}\\
		g_{V\pi\pi}&=&\int_0^{z_m} dz\frac{e^{-\phi(z)}}{z^3}\psi_{V^{(n)}}^{a}\psi_{\pi^{(m)}}^b\psi_{\pi^{(k)}}^c(h^{abc}+h^{acb}-2g^{cab}).\label{3cc-4}
	\end{eqnarray}
	where $\psi$ is the eigenfunction of the mesons.
	
	Similar to the three-point interaction of mesons, the four-point interaction can be obtained from the quartic order term of 5D action. 
	From Eq.(\ref{eom-4}), it can be seen that the quartic order of the action can be divided into seven parts $S_{VVVV}$, $S_{VVAA}$, $S_{AAAA}$, $S_{VV\pi\pi}$, $S_{AA\pi\pi}$, $S_{A\pi\pi\pi}$ and $S_{\pi\pi\pi\pi}$, where	
	\begin{eqnarray}
		S_{VVVV}&=&-\int_0^{z_m} d^5x\frac{e^{-\phi(z)}}{4g_5^2z}f^{abcd}V^{\mu,a}V^{\nu,b}V_\mu^cV_\nu^d,\\
		S_{VVAA}&=&-\int_0^{z_m} d^5x\frac{e^{-\phi(z)}}{4g_5^2z}\Bigg\{2V^{\mu,a}V^{\nu,b}A_\mu^cA_\nu^d(f^{abcd}+f^{cbad})+2V^{\mu,a}V_\mu^{b}A^{\nu,c}A_\nu^df^{acbd}\Bigg\},\\
		S_{AAAA}&=&-\int_0^{z_m} d^5x\frac{e^{-\phi(z)}}{4g_5^2z}f^{abcd}A^{\mu,a}A^{\nu,b}A_\mu^cA_\nu^d,\\
		S_{VV\pi\pi}&=&-\int_0^{z_m} d^5x\frac{e^{-\phi(z)}}{z^3}V^{\mu,a}V_\nu^b\pi^c\pi^d(h^{abcd}-g^{acbd}),\\
		S_{AA\pi\pi}&=&-\int_0^{z_m} d^5x\frac{e^{-\phi(z)}}{z^3}A^{\mu,a}A_\mu^b\pi^c\pi^d(k^{acbd}-l^{abcd}),\\
		S_{A\pi\pi\pi}&=&-\int_0^{z_m} d^5x\frac{e^{-\phi(z)}}{z^3}A^{\mu,a}\partial_\mu\pi^b\pi^c\pi^d(l^{bacd}+\frac{l^{abcd}}{3}+\frac{l^{acbd}}{3}+\frac{l^{acdb}}{3}-k^{bcad}-k^{cbad}),\\
		S_{\pi\pi\pi\pi}&=&-\int_0^{z_m} d^5x\frac{e^{-\phi(z)}}{z^3}(\partial^\mu\pi^a\partial_\mu\pi^b\pi^c\pi^d+\partial_z\pi^a\partial_z\pi^b\pi^c\pi^d)(\frac{k^{acbd}+k^{cabd}+k^{acdb}+k^{cadb}}{4}\nonumber\\
		&&-\frac{l^{abcd}+l^{acbd}+l^{acdb}}{3}).
	\end{eqnarray}

    Similarly, by substituting the eigenfunctions into the action, the four-point coupling constants can be obtained as

    \begin{eqnarray}
    	g_{VVVV}&=&\int_0^{z_m} dz\frac{e^{-\phi(z)}}{4g_5^2z}f^{abcd}\psi_{V^{(n)}}^a\psi_{V^{(m)}}^b\psi_{V^{(k)}}^c\psi_{V^{(j)}}^d,\label{4cc-1}\\
    	g_{VVAA}&=&\int_0^{z_m} dz\frac{e^{-\phi(z)}}{4g_5^2z}2\psi_{V^{(n)}}^a\psi_{V^{(m)}}^b\psi_{A^{(k)}}^c\psi_{A^{(j)}}^d(f^{abcd}+f^{cbad}),\label{4cc-2}\\
    	g_{AAAA}&=&\int_0^{z_m} dz\frac{e^{-\phi(z)}}{4g_5^2z}f^{abcd}\psi_{A^{(n)}}^a\psi_{A^{(m)}}^b\psi_{A^{(k)}}^c\psi_{A^{(j)}}^d,\label{4cc-3}\\
    	g_{VV\pi\pi}&=&\int_0^{z_m} dz\frac{e^{-\phi(z)}}{z^3}\psi_{V^{(n)}}^a\psi_{V^{(m)}}^b\psi_{\pi^{(k)}}^c\psi_{\pi^{(j)}}^d(h^{abcd}-g^{acbd}),\label{4cc-4}\\
    	g_{AA\pi\pi}&=&\int_0^{z_m} dz\frac{e^{-\phi(z)}}{z^3}\psi_{A^{(n)}}^a\psi_{A^{(m)}}^b\psi_{\pi^{(k)}}^c\psi_{\pi^{(j)}}^d(k^{acbd}-l^{abcd}),\label{4cc-5}\\
    	g_{A\pi\pi\pi}&=&\int_0^{z_m} dz\frac{e^{-\phi(z)}}{z^3}\psi_{A^{(n)}}^a\psi_{\pi^{(m)}}^b\psi_{\pi^{(k)}}^c\psi_{\pi^{(j)}}^d(l^{bacd}+\frac{l^{abcd}}{3}+\frac{l^{acbd}}{3}+\frac{l^{acdb}}{3}-k^{bcad}-k^{cbad}),\label{4cc-6}\\
    	g_{\pi\pi\pi\pi}&=&\int_0^{z_m} dz\frac{e^{-\phi(z)}}{z^3}(\psi_{\pi^{(n)}}^a\psi_{\pi^{(m)}}^b\psi_{\pi^{(k)}}^c\psi_{\pi^{(j)}}^d+\psi_{\pi^{(n)}}^{\prime,a}\psi_{\pi^{(m)}}^{\prime,b}\psi_{\pi^{(k)}}^c\psi_{\pi^{(j)}}^d)(\frac{k^{acbd}+k^{cabd}+k^{acdb}+k^{cadb}}{4}\nonumber\\
    	&&-\frac{l^{abcd}+l^{acbd}+l^{acdb}}{3}),\label{4cc-7}
    \end{eqnarray}
    where $\prime$ represents the derivative with respect to $z$.
	
	\section{Numerical analysis}\label{sec-5}
	
	In this section, we present the numerical results for the spectra, decay constants and coupling constants. 
	In our holographic model, there are six physical parameters: quadratic term coefficient $\mu$ of dilaton field, IR cutoff $z_m$, constants of vacuum expectation values $C_1(q)$ for $q=(l, s, c)$ and constant of auxiliary field $D_2$ . These parameters are fixed by the following meson masses, as specified in the Table \ref{fitdata}.
	
	\begin{table}[th]
		\caption{Experimental data of meson masses were used to fit the parameters $\mu$, $z_m$, $D_1$ and $C_1(q)$. The experimental data are taken from \cite{ParticleDataGroup:2020ssz}.} \label{fitdata}
		\begin{tabular}{|c|c|c|c|c||c|c|c|c|c|}
			\hline
			Resonance & Quark content & $J^{P}$ & Exp.(MeV) & Model(MeV) & Resonance & Quark content & $J^{P}$ & Exp.(MeV) & Model(MeV)\\
			\hline
			$\rho^{0}$&$\bar{u}u,\bar{d}d$&$1^-$&$775$&$860$&$a_{1}^{-}$&$\bar{u}d$&$1^+$&$1230$&$1222$\\
			\hline
			$K^{*}$&$\bar{d}s,\bar{s}d$&$1^-$&$892$&$884$&$\chi_{c1}$&$\bar{c}c$&$1^+$&$3511$&$3464$\\
			\hline
			$J/\psi$&$\bar{c}c$&$1^-$&$3097$&$3098$&$\psi(3770)$&$\bar{c}c$&$1^-$&$3773$&$3712$\\
			\hline
		\end{tabular}
	\end{table}
	Since the equation of motion of the vector meson $\rho$ has ${M_V^a}^2 =0$, it is used to fit $\mu$. The parameter $\mu$ is chosen to be $0.43$GeV, in which case both the mass and the Regge slope of $\rho$ meson can be fitted. 
	After $\mu$ is chosen, the masses and Regge slope of the mesons $a_{1}^{-}$ ,$K^{_{*-}}$ and $\chi_{c1}$ are used to fix $C_1(l)$, $C_1(s)$ and $C_1(c)$.  
	The remaining two parameters $z_m$ and $D_1$ are fixed by the mass of $J/\psi$ and its Regge slope, where the Regge slope is replaced with the mass of $\psi(3770)$. Thus, the parameters are finally fixed by the mass of $J/\psi$ and $\psi(3770)$.
	Through the expansion of the vacuum expectation value $v_{l,s,c}$ at UV boundary, the parameters $C_1(l)$, $C_1(s)$ and $C_1(c)$ can be translated into quark masses $M_{l,s,c}$ and quark condensation $\Sigma_{l,s,c}$. 
	With the numerical fitting strategy described above, the quark masses and condensation are chosen as: $M_{l}=140~$MeV, $M_{s}=200~$MeV, $M_{c}=1200~$MeV and $\Sigma_{l}=(135~\rm{MeV})^{3}$, $\Sigma_{s}=(152~\rm{MeV})^{3}$, $\Sigma_{c}=(276~\rm{MeV})^{3}$.
	Similarly, the parameter $D_1$ can also be translated into quark mass and condensation as $m_c=1020~$MeV and $\sigma_{c}=(262~\rm{MeV})^{3}$.
	
\begin{figure}[!thb]
    	\centering
    	\includegraphics[width=0.8\textwidth]{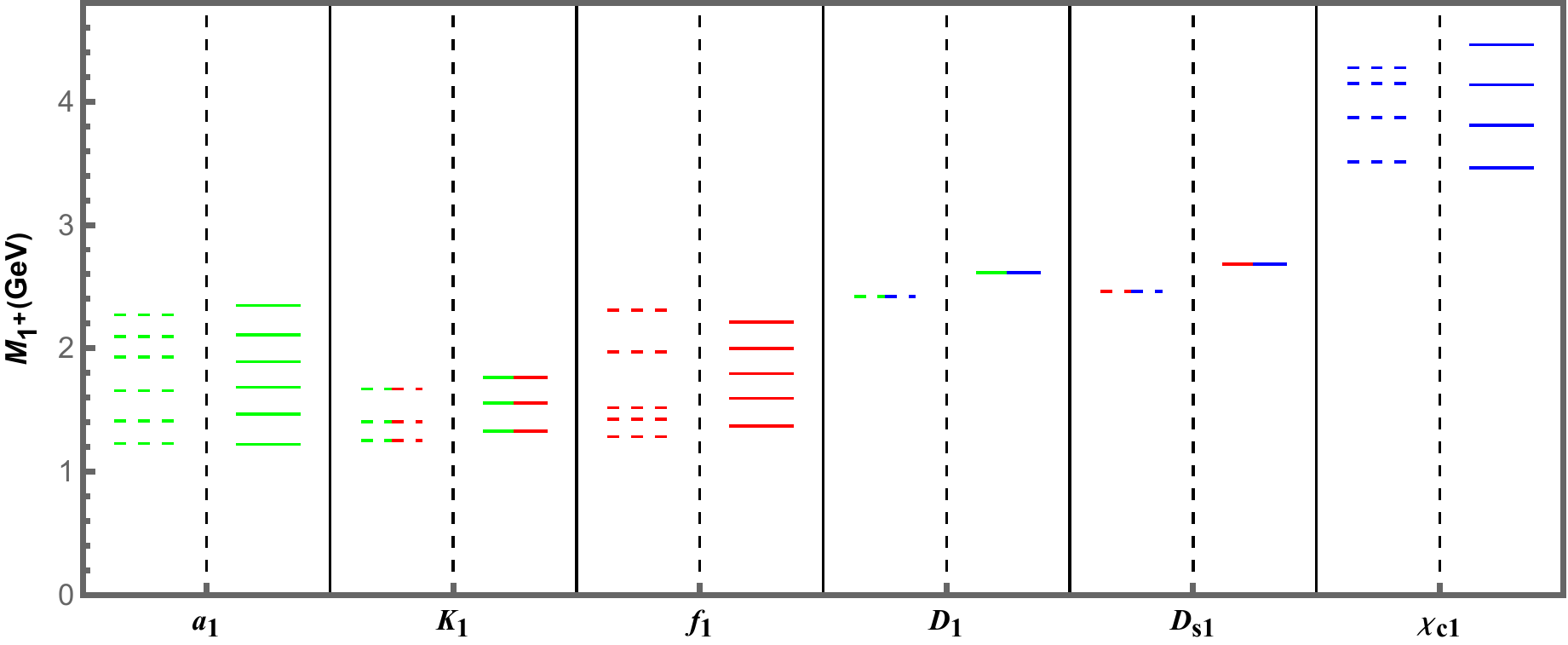}
    	\includegraphics[width=0.8\textwidth]{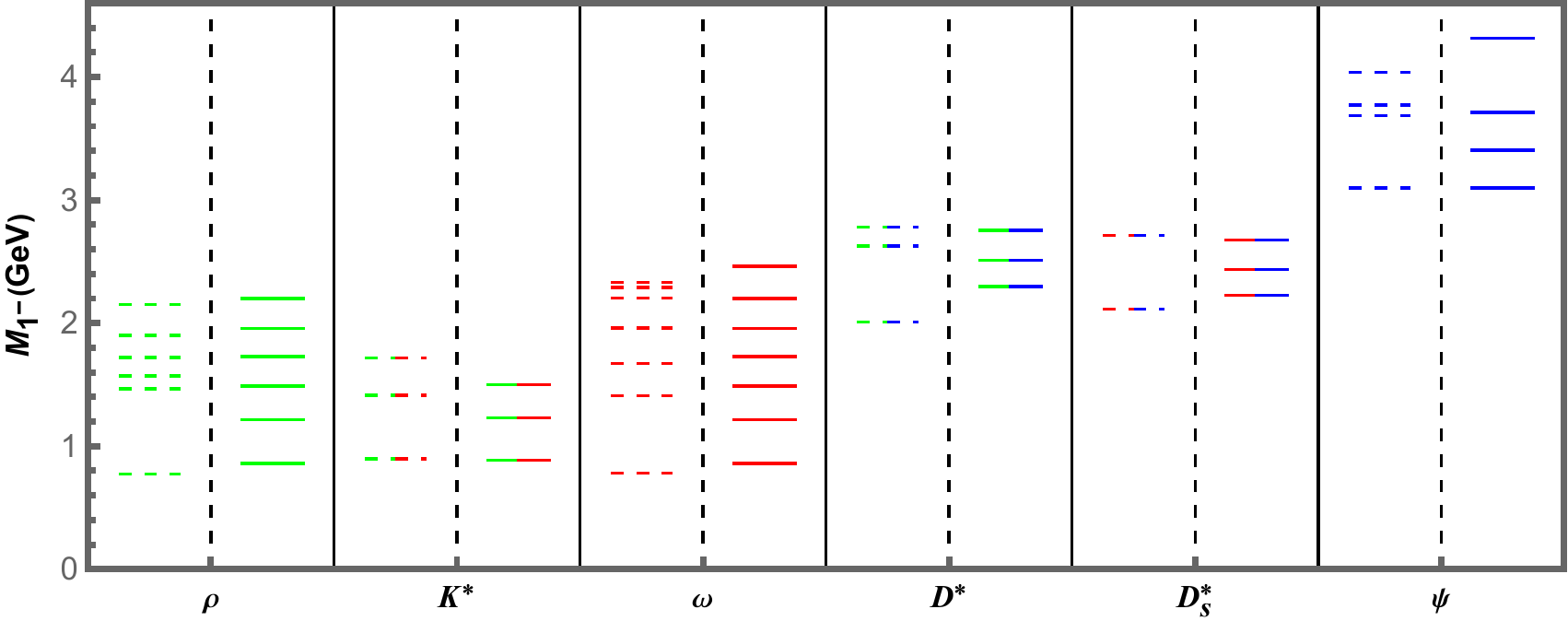}
    	\includegraphics[width=0.8\textwidth]{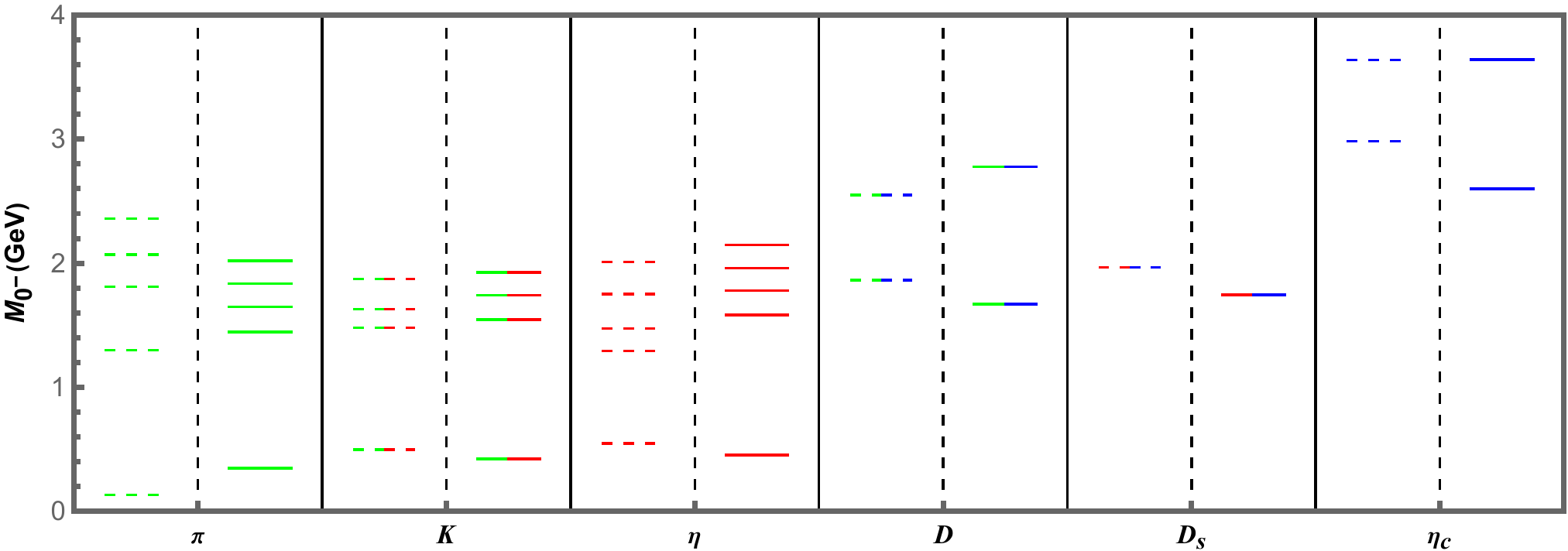}
    	\caption{The meson spectra for $N_f=2+1+1$, where the dashed and solid lines represent the experimental data and model results, respectively. Different colors indicate different quarks, where $ud$ quark, $s$ quark, and $c$ quark are indicated by green, red, and blue, respectively.}
    	\label{fig:mesonspectra}	
    \end{figure}

	By fixing the parameters of the model, the masses of pseudoscalar, vector and axial vector mesons and their excited states can be obtained. The model predictions and experimental data of the mesons masses are listed in Table \ref{spectra}, where the experimental data are taken from \cite{ParticleDataGroup:2020ssz}. Figure \ref{fig:mesonspectra} shows the graphical display of Table \ref{spectra}.
	
	\begin{table}[th]
		\caption{The masses of the mesons are predicted by the model. The experimental data are taken from \cite{ParticleDataGroup:2020ssz}.} \label{spectra}
		\begin{tabular}{|c||c|c|c||c|c|c||c|c|c|}
			\hline
			Quark content & $0^{-}$ & Exp.(MeV) & Mod.(MeV)& $1^{-}$ & Exp.(MeV) & Mod.(MeV)& $1^{+}$ & Exp.(MeV) & Mod.(MeV)\\
			\hline
			$\bar{u}u, \bar{d}d, \bar{s}s$&$\pi^{0}$&$135$&$349$&$\rho(770)$&$775$&$860$&$a_1(1260)$&$1230$&$1222$\\
			\hline
			&$\pi(1300)$&$1300$&$1446$&$\rho(1450)$&$1465$&$1216$&$a_1(1420)$&$1411$&$1468$\\
			\hline
			&$\pi(1800)$&$1810$&$1649$&$\rho(1570)$&$1570$&$1490$&$a_1(1640)$&$1655$&$1685$\\
			\hline
			&$\pi(2070)$&$2070$&$1835$&$\rho(1700)$&$1720$&$1727$&$a_1(1930)$&$1930$&$1892$\\
			\hline
			&$\pi(2360)$&$2360$&$2021$&$\rho(1900)$&$1900$&$1957$&$a_1(2095)$&$2096$&$2111$\\
			\hline
			&&&&$\rho(2150)$&$2150$&$2202$&$a_1(2270)$&$2270$&$2346$\\
			\hline
			$\bar{s}d, \bar{d}s$&$K^{0}$&$498$&$424$&$K^{*}(892)$&$892$&$884$&$K_1(1270)$&$1253$&$1328$\\
			\hline
			&$K(1460)$&$1482$&$1546$&$K^{*}(1410)$&$1414$&$1230$&$K_1(1400)$&$1403$&$1557$\\
			\hline
			&$K(1630)$&$1629$&$1743$&$K^{*}(1680)$&$1718$&$1500$&$K_1(1650)$&$1672$&$1763$\\
			\hline
			&$K(1830)$&$1874$&$1925$&&&&&&\\
			\hline
			$\bar{u}u, \bar{d}d, \bar{s}s$&$\eta$&$548$&$454$&$\omega(782)$&$782$&$860$&$f_1(1285)$&$1282$&$1369$\\
			\hline
			&$\eta(1295)$&$1294$&$1585$&$\omega(1420)$&$1410$&$1216$&$f_1(1420)$&$1426$&$1593$\\
			\hline
			&$\eta(1475)$&$1475$&$1779$&$\omega(1650)$&$1670$&$1490$&$f_1(1510)$&$1518$&$1796$\\
			\hline
			&$\eta(1760)$&$1751$&$1962$&$\omega(1960)$&$1960$&$1727$&$f_1(1970)$&$1971$&$1998$\\
			\hline
			&$\eta(2010)$&$2010$&$2148$&$\omega(2205)$&$2205$&$1957$&$f_1(2310)$&$2310$&$2214$\\
			\hline
			&&&&$\omega(2290)$&$2290$&$2202$&&&\\
			\hline
			&&&&$\omega(2330)$&$2330$&$2462$&&&\\
			\hline
			$\bar{c}u, \bar{u}c$&$D^{0}$&$1865$&$1671$&$D^{*}(2007)^0$&$2007$&$2296$&$D_1(2420)$&$2422$&$2615$\\
			\hline
			&$D_{0}(2550)^0$&$2549$&$2778$&$D_1^{*}(2600)^0$&$2627$&$2512$&&&\\
			\hline
			&&&&$D_1^{*}(2760)^0$&$2781$&$2756$&&&\\
			\hline
			$\bar{c}s, \bar{s}c$&$D_s^{\pm}$&$1968$&$1746$&$D_{s1}^{*\pm}$&$2112$&$2227$&$D_{s1}(2460)$&$2460$&$2682$\\
			\hline
			&&&&$?$&$?$&$2436$&&&\\
			\hline
			&&&&$D_{s1}^{*}(2700)^{\pm}$&$2714$&$2674$&&&\\
			\hline
			$\bar{c}c$&$\eta_c$&$2984$&$2600$&$J/\psi(1S)$&$3097$&$3098$&$\chi_{c1}(1P)$&$3511$&$3464$\\
			\hline
			&$\eta_c(2S)$&$3637$&$3641$&$\psi(2S)$&$3686$&$3773$&$\chi_{c1}(3872)$&$3872$&$3808$\\
			\hline
			&&&&$\psi(3770)$&$3773$&$3712$&$\chi_{c1}(4140)$&$4147$&$4138$\\
			\hline
			&&&&$\psi(4040)$&$4040$&$4317$&$\chi_{c1}(4274)$&$4274$&$4460$\\
			\hline
		\end{tabular}
	\end{table}

    As can be seen in Fig.\ref{fig:mesonspectra}, for axial vector mesons, the $a_1$ and $K_1$ mesons results are in good agreement with the data. For the $f_1$ meson, the results are slightly deviated from the data because the mixing of $s$ and $u, d$ quarks are not considered. For $D_1$ and $D_{s1}$ mesons, the model calculations are about $0.2$GeV heavier than the data. For $\chi_{c1}$ mesons, the ground state and the first two excited states fit the experimental data relatively well, while the third excited state is about $0.2$GeV heavier than the data. 
    
    For vector mesons, the calculations are degenerate because the model cannot distinguish between $\rho$ and $\omega$ mesons. From the results, it can be seen that the $\rho$ meson fits better compared to the $\omega$. This is also understandable due to the mixing of $s$ quark with $u, d$ quarks being neglected. For the $K^{*}$ meson, its excited state results are about $0.2$GeV lighter than the experimental data, so it can be seen that its Regge behavior does not fit very well. The reason for this result could be that the mass of the $s$ quark fitted by the model is too close to the mass of the $u, d$ quark. For the $D^{*}$ meson, the model gives the ground state mass that is about $0.3$GeV heavier than the data, while the excited state fits relatively well. For $\psi$ mesons, the first and third excited states do not fit very well compared to the data.
    
    It is worth noting that only the excited state $D_{s1}^*(2700)$ was found in the experiment, which is close to the mass of the second excited state of $D_{s1}$ in the model. Therefore, our model predicts the possible existence of a new excited state with a mass of roughly $2436~$MeV.
    
    For pseudoscalar mesons, the second, third and fourth excited states of $\pi$ mesons are about $0.15-0.3$GeV lighter compared to the data, so the Regge slope of $\pi$ is not well matched. From the data, it can be found that the Regge slope of $\pi$ meson is higher than that of other light mesons, so it is challenging to solve this problem in the holographic model. For $\eta$ mesons, the calculated excited state masses are about $0.25$GeV heavier compared to the data, but their Regge behavior remains the same. The reason for this may also come from the absence of the mixing term. For $D$ and $D_s$ mesons, their ground state calculations are $0.2$GeV lighter compared to the data, while the ground state results for $\eta_c$ mesons are $0.4$GeV lighter.    
	
\begin{figure}[!thb]
		\centering
		\includegraphics[width=0.32\textwidth]{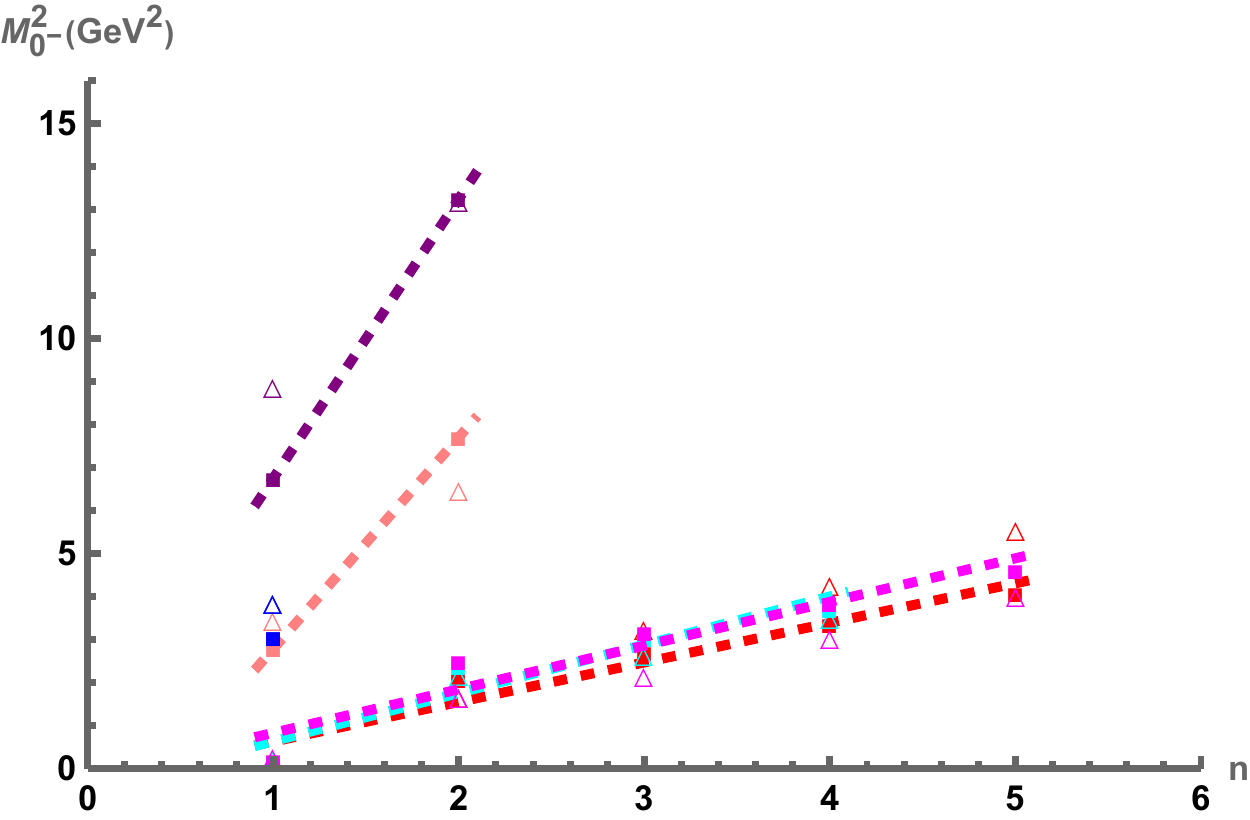}
		\includegraphics[width=0.32\textwidth]{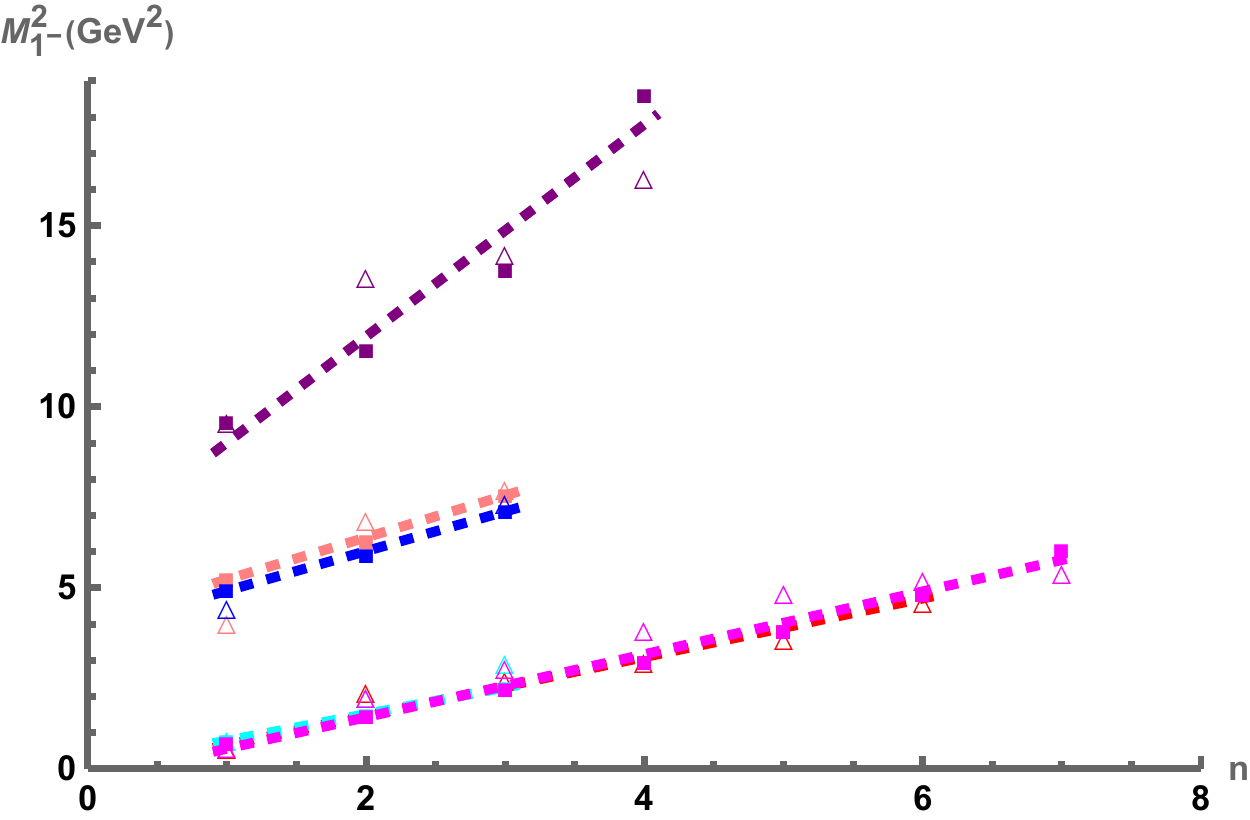}
		\includegraphics[width=0.32\textwidth]{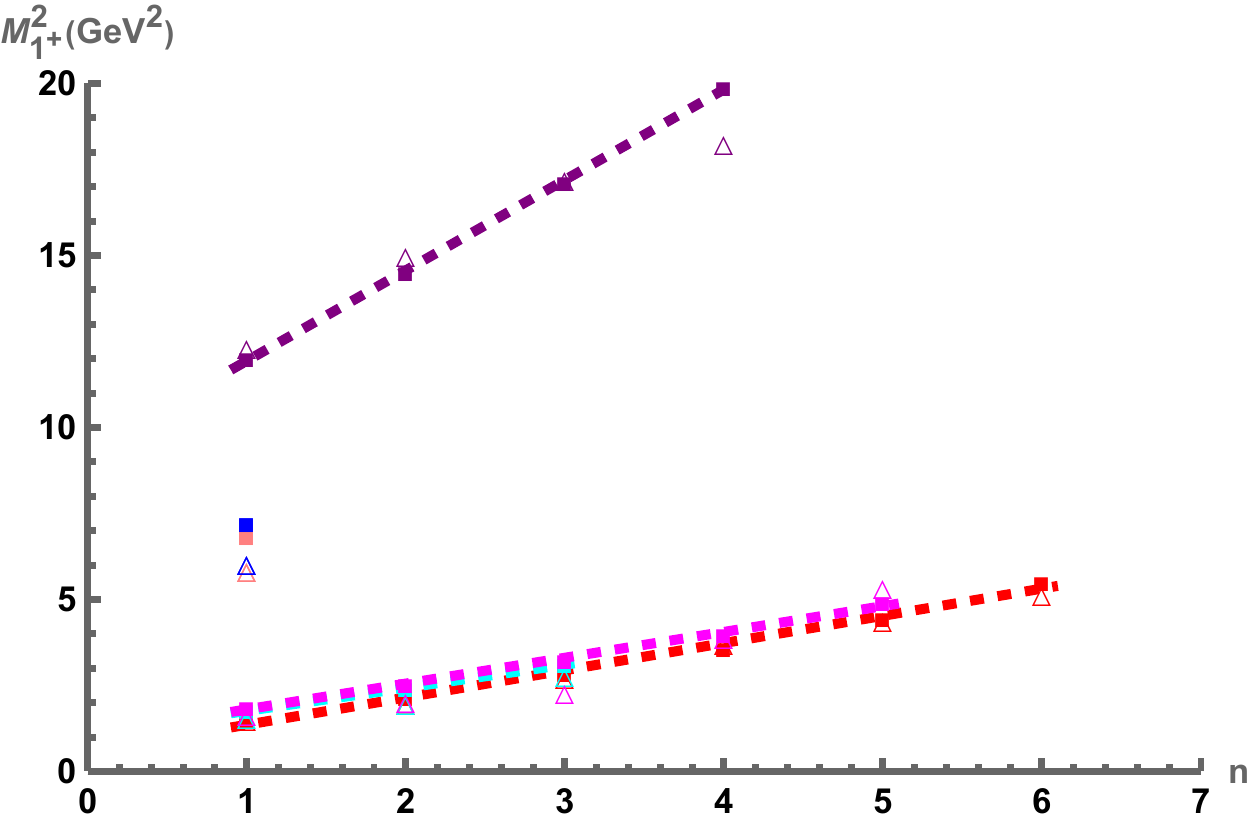}
		\caption{Comparison of light meson and charm meson Regge slopes. Where the hollow triangles and squares represent the experimental data and model results, respectively. The dashed line is the linear fit of the model results. The different colors indicate the different components of the mesons, specifically selected as: red($\bar{l}l$), cyan($S=\pm 1$), magenta($\bar{s}s$), pink($C=\pm 1$), blue($C=\pm 1$, $S=\pm 1$), purple($\bar{c}c$).}\label{fig:slope}
	\end{figure}

	Due to the introduction of IR cutoff $z_m$ and additional auxiliary field $H$, different Regge trajectories for light and heavy mesons can be achieved. Among them, the $H$ field serves to improve the intercept of the Regge behavior, while the IR cutoff $z_m$ can increase the slope of the heavy mesons. The Regge trajectories of light mesons and charm mesons can be seen in Fig.\ref{fig:slope}, where different colors represent different components of mesons.
	
    The model predicts the decay constants of the mesons, and their ratios are shown in Table \ref{mesondecay}. 
    The measurements are also shown in the table, where the decay constants of pseudoscalar mesons are from Ref.\cite{ParticleDataGroup:2020ssz,FlavourLatticeAveragingGroup:2019iem,McNeile:2012qf}, vector and axial vector mesons from Ref.\cite{Ballon-Bayona:2017bwk}.
    
    \begin{table}[th]
    	\caption{The ratio of decay constants predicted by the model. The measured
    		value are taken from \cite{ParticleDataGroup:2020ssz,FlavourLatticeAveragingGroup:2019iem,McNeile:2012qf, Ballon-Bayona:2017bwk}} \label{mesondecay}
    	\begin{tabular}{|c|c|c||c|c|c||c|c|c|}
    		\hline
    		Observable  &Exp./LQCD&Model&Observable&Ref.\cite{Ballon-Bayona:2017bwk}&Model
    		&Observable&Ref.\cite{Ballon-Bayona:2017bwk}&Model\\
    		\hline
    		&&&&&&&&\\
    		$f_{K}/f_{\pi}$&$1.196$&$1.199$&$f_{K^*}^{1/2}/f_{\rho}^{1/2}$&$1.005$&$0.973$
    		&$f_{K_1}^{1/2}/f_{a_1}^{1/2}$&$1.085$&$0.706$\\
    		&&&&&&&&\\
    		$f_{D_s}/f_{D}$&$1.180$&$1.040$&$f_{D_s^*}^{1/2}/f_{D^*}^{1/2}$&$0.954$&$1.951$
    		&$f_{D_{s1}}^{1/2}/f_{D_1}^{1/2}$&&$0.504$\\
    		&&&&&&&&\\
    		$f_{\eta_c}/f_{D_s}$&$1.576$&$1.427$&&&&&&\\
    		&&&&&&&&\\
    		\hline
    	\end{tabular}
    \end{table}

    As can be seen from the table \ref{mesondecay}, for pseudoscalar mesons, the ratios of decay constants are close to the data. Among them, the ratio $f_{K}/f_{\pi}$ is in good agreement with the data, while the results of ratios $f_{D_s}/f_{D}$ and $f_{\eta_c}/f_{D_s}$ are smaller than the data by about $0.1$.
    
    For the vector and axial vector decay constants, the model results are compared with Ref.\cite{Ballon-Bayona:2017bwk}. It can be seen that for $f_{K^*}^{1/2}/f_{\rho}^{1/2}$, $f_{D_s^*}^{1/2}/f_{D^*}^{1/2}$, and $f_{K_1}^{1/2}/f_{a_1}^{1/2}$, the model results are opposite to the Ref.\cite{Ballon-Bayona:2017bwk} results. The ratio greater than $1$ in our model is less than $1$ in Ref.\cite{Ballon-Bayona:2017bwk}, and vice versa. For this result, it may come from the introduction of the dilaton field, which deform the configuration of the 5D wave function, generating Regge trajectories while reversing the ratio of the decay constants. In addition, the ratio $f_{D_{s1}}^{1/2}/f_{D_1}^{1/2}$ is also predicted by the model.
    
    In the model, coupling constants for cubic and quartic vertices can be obtained by substituting the eigenfunctions of the mesons into Eq.(\ref{3cc-1}-\ref{3cc-4}) and Eq.(\ref{4cc-1}-\ref{4cc-7}). The ratios of the coupling constants $(D^{(*)}, D, A)$, $(D^{(*)}, D^{(*)}, V)$, $(D^{(*)}, D^{(*)}, V)$ and $(\psi, D^{(*)}, D, P)$ are calculated in the model and their results are listed in Table \ref{meson-cc}.

	\begin{table}[!th]
		\caption{Coupling constants for model calculations.}
		\label{meson-cc} 
		\begin{tabular}{|c|c||c|c|}
			\hline
			Observable&Model&Observable&Model\\
			\hline
			&&&\\
			$g_{K^* D^* D_s^*}/g_{\rho D^* D^*}$&$1.038$&$g_{K^* D D_s}/g_{\rho D D}$&$0.203$\\
			&&&\\
			\hline
			&&&\\
			$g_{K_1 D_s D^*}/g_{a_1 D D^*}$&$0.433$&$g_{\psi D_s^* D K}/g_{\psi D^* D \pi}$&$0.435$\\
			&&&\\
			\hline
		\end{tabular}
	\end{table}

    \section{Conclusion}
    \label{sec-conclusion}
    In summary, in this paper, the soft wall model with $N_f=2$\cite{Karch:2006pv} is extended to the case of $N_f=4$, where a light scalar field $X$ and a heavy scalar field $H$ are introduced, separately. The $H$ field is responsible for the breaking of $SU(N_f=4)$ to $SU(N_f=3)$.  The ground state and its Regge excitation of meson spectra in the light flavor sector and heavy flavor sector as well as the ligh-heavy mesons are well in agreement with the Particle data group (PDG). Due to the additional introduction of the $H$ field in the model, different Regge slopes for light and heavy mesons can be achieved. The $N_f=4$ holographic model consists of six parameters including the Regge slope $\mu$, quark masses $C_1(q)$ with $q=(l, s, c)$ and $D_1$. These parameters are fitted to the experimental masses of $\rho$, $a_{1}$, $K^*$, $\chi_{c1}$, $J/\psi$
    and $\psi(3770)$ mesons. 
    The masses of other pseudoscalars, vectors and axial vector mesons as well as the ratio of decay constants are calculated in the model and compared with the experimental data.
    In addition to this, the coupling constants $(D^{(*)}, D, A)$, $(D^{(*)}, D^{(*)}, V)$, $(D^{(*)}, D^{(*)}, V)$ and $(\psi, D^{(*)}, D, P)$ are also estimated in the model.
	
	\begin{acknowledgements}
		This work is supported by the NSFC under Grant Nos. 11725523 and 11735007,  Chinese Academy of Sciences under Grant No. XDPB09, the start-up funding from University of Chinese Academy of Sciences(UCAS), and the Fundamental Research Funds for the Central Universities.
	\end{acknowledgements}

\end{document}